\documentclass[aps,prl,twocolumn]{revtex4}

\usepackage{amsmath, mathtools,amssymb}    
\usepackage{graphicx}  
\usepackage{natbib}  
\usepackage{hyperref}
\usepackage{epstopdf}
\usepackage{subfigure}
\usepackage{pstricks}
\usepackage{color}
\usepackage{mathrsfs}

\newcommand{\nn}{\nonumber \\}

\begin{document}

\title{\bf Strange metal state near a heavy-fermion quantum critical point}   
\author{Yung-Yeh, Chang$^{1}$, Silke Paschen$^{2}$, and Chung-Hou Chung$^{1,3}$}
 \affiliation{
$^{1}$Department of Electrophysics, National Chiao-Tung University, 1001 University Rd., Hsinchu, 300 Taiwan, R.O.C.\\
$^{2}$Institute of Solid State Physics, TU Vienna, Wiedner Hauptstrasse 8-10, 1040 Vienna, Austria\\
$^{3}$Physics Division, National Center for Theoretical Sciences, HsinChu, 300 Taiwan, R.O.C.}
 \email{cdshjtr.ep02g@nctu.edu.tw}
 \email{chung@mail.nctu.edu.tw}
 \email{paschen@ifp.tuwien.ac.at} 
  \date{\today}

   \begin{abstract}    
Recent experiments on quantum criticality in the Ge-substituted heavy-electron material YbRh$_2$Si$_2$ under magnetic field have revealed a possible 
non-Fermi liquid (NFL) strange metal (SM) state over a finite range of fields at low temperatures, which still remains a puzzle. In the SM region, the zero-field antiferromagnetism is suppressed. Above a critical field, it gives way to a heavy Fermi liquid with Kondo correlation. The $T$ (temperature)-linear resistivity and the $T$-logarithmic followed by a power-law singularity in the specific heat coefficient 
at low $T$, salient NFL behaviours in the SM region, are un-explained. 
We offer a mechanism to address these open issues theoretically based on the competition between a quasi-$2d$ fluctuating 
 short-ranged resonant-valence-bonds (RVB) spin-liquid and the Kondo correlation near criticality.  
Via a field-theoretical renormalization group analysis on an effective field theory beyond a large-$N$   
approach to an antiferromagnetic Kondo-Heisenberg model, we identify the critical point, and 
explain remarkably well both the crossovers 
and the SM behaviour. 
 
   \end{abstract}
     \maketitle
        {\it Introduction.}
Magnetic field tuned quantum phase transitions (QPTs)\cite{SachdevQPT} in heavy-fermion metals of both pure and Ge-substituted $\text{YbRh}_2\text{Si}_2$ (YRS) compounds \cite{Gegenwart2008, Custers2003, Grosche2001a, Custers2010, Coleman2007} are of great interest both theoretically and experimentally.
Near a quantum critical point (QCP), these systems show exotic non-Fermi liquid (NFL) electronic properties at finite temperatures, including a $T$-linear resistivity \cite{Custers2010, SiNature2008, Custers2003} and power law-in-$T$ at low $T$ followed by a $T$-logarithmic specific heat coefficient at higher $T$ \cite{Custers2003}, which still remain as outstanding open issues \cite{QMSi2001, Zhu2002, Si2003, Si2011,Senthil2003,Wolfle2011, Abrahams2012,Rosch1997,2005Pepin}. 

For the QPT in pure YRS, the ``Kondo breakdown" scenario \cite{QMSi2001} offers a general understanding: competition between the Kondo 
and antiferromagnetic RKKY couplings 
leads to a QCP, separating the antiferromagnetic metallic state from the 
paramagnetic Landau Fermi-liquid state (LFL) with enhanced Kondo correlation. 
There, 
a magnetic phase transition and the emergence (or the breakdown) of the Kondo effect occur simultaneously and in the ground state the system  undergoes a jump from a small to a large Fermi surface \cite{PaschenNat2004, FriedemannPNAS}.  
The understanding of the NFL properties near the QCP still remains an outstanding open issue though certain aspects have been addressed \cite{QMSi2001, Zhu2002, Si2003, Si2011,Senthil2003,Senthil2004,Wolfle2011, Abrahams2012,Rosch1997,2005Pepin}. 
    
\begin{figure*}[ht]
                  \includegraphics[width=\textwidth]{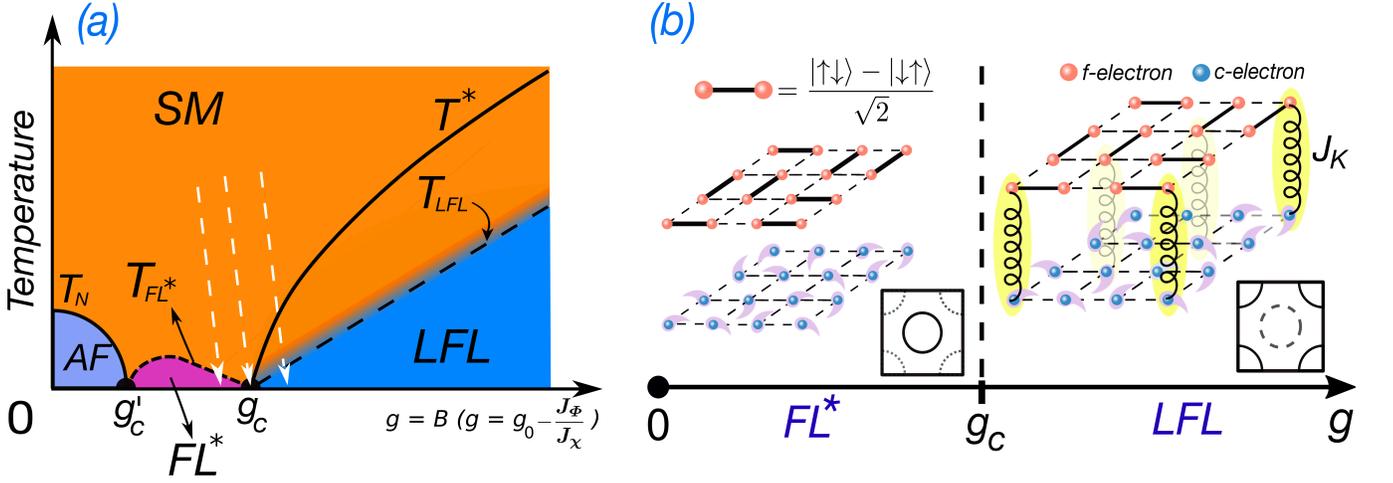}
\vskip -0.2cm
                  \caption{(a) Schematic phase diagram of Ge-substituted YRS \cite{Custers2003, Custers2010}. The horizontal axis refers to either the applied magnetic field $B$ or $g=g_0-J_\Phi/J_\chi$ with $J_\Phi=J_H$, $J_\chi=J_K$. The solid curve $T^\ast$ labels the small-to-large Fermi surface crossover (see (b)), while $T_{LFL}$ ($T_{FL^\ast}$) refers to the crossover scale from quantum critical region to LFL (FL$^\ast$), respectively. 
 $g_{c}$ and $g_{c}'$ denote the quantum critical points for AF-FL$^\ast$ and FL$^\ast$-LFL transitions, respectively. 
 White dashed arrows refer to the RG flows. (b) Schematic plot describing the competition between the RVB spin-liquid FL$^\ast$ phase and the Kondo LFL phase near $g_c$ at zero temperature. 
 The black heavy lines between two adjacent $f$-electrons refer to the RVB spin-singlets, while the helical lines embedded in yellow stand for the Kondo hybridization.} 
                  \label{fig:phasediag}
   \end{figure*}

Recent experiments on Ge-substituted YRS, however, reveals intriguing distinct features. 
 First, the magnetic phase transition at $g_c^\prime$ 
occurs at a lower field than the Kondo destruction  at $g_c$, 
leading to a decoupling of the AFM and the LFL phases (see Fig. \ref{fig:phasediag}(a)) \cite{Custers2003}.
  Interestingly, similar NFL behaviour persists over a finite range in magnetic fields at the lowest temperatures, suggesting a possible exotic novel stable spin-disordered 
``strange-metal'' (SM) ground state \cite{Custers2010}. 
Open questions to be addressed include: 
  Dose this SM behaviours come from the SM ground state or from a single QCP?
  What is the role played by the magnetic field? What mechanism is behind the Kondo breakdown at the QCP? 
 
A short-ranged resonanting-valence-bond (RVB) spin-liquid (SL) picture has recently been proposed to describe the metallic spin-liquid phase of heavy fermion metal with frustrated antiferromagnetic RKKY coupling \cite{Pixley2013}. A generic phase diagram was proposed in Ref. \onlinecite{Pixley2013} in terms of magnetic frustration and Kondo correlation. There, a small Kondo coupling at low fields may co-exist with a small Fermi surface to form a (Kondo stablized) SL metal \cite{ColemanAndrei1989}.  The disorder due to Ge substitution here makes this proposal more attractive to account for 
the SM state. 
It is of great interest to further explore the mechanism of this behaviour. 
 In this work, we propose a mechanism for the Kondo breakdown and the quantum criticality in Ge-substituted YRS  at $g_c$ via a fermionic large-N approach based on the symplectic group $Sp(N)$ symmetry on a quasi-$2d$ Kondo lattice. 
The competition between the 
Gaussian fluctuating RVB spin-liquid and the Kondo correlation (see Fig.~1) explains remarkably well both the transition and the NFL behaviours. 
The SM state indicated in experiments can be understood 
as quantum critical region near critical Kondo breakdown. 

\begin{figure}[ht]
              \centering
                \includegraphics[scale=0.3]{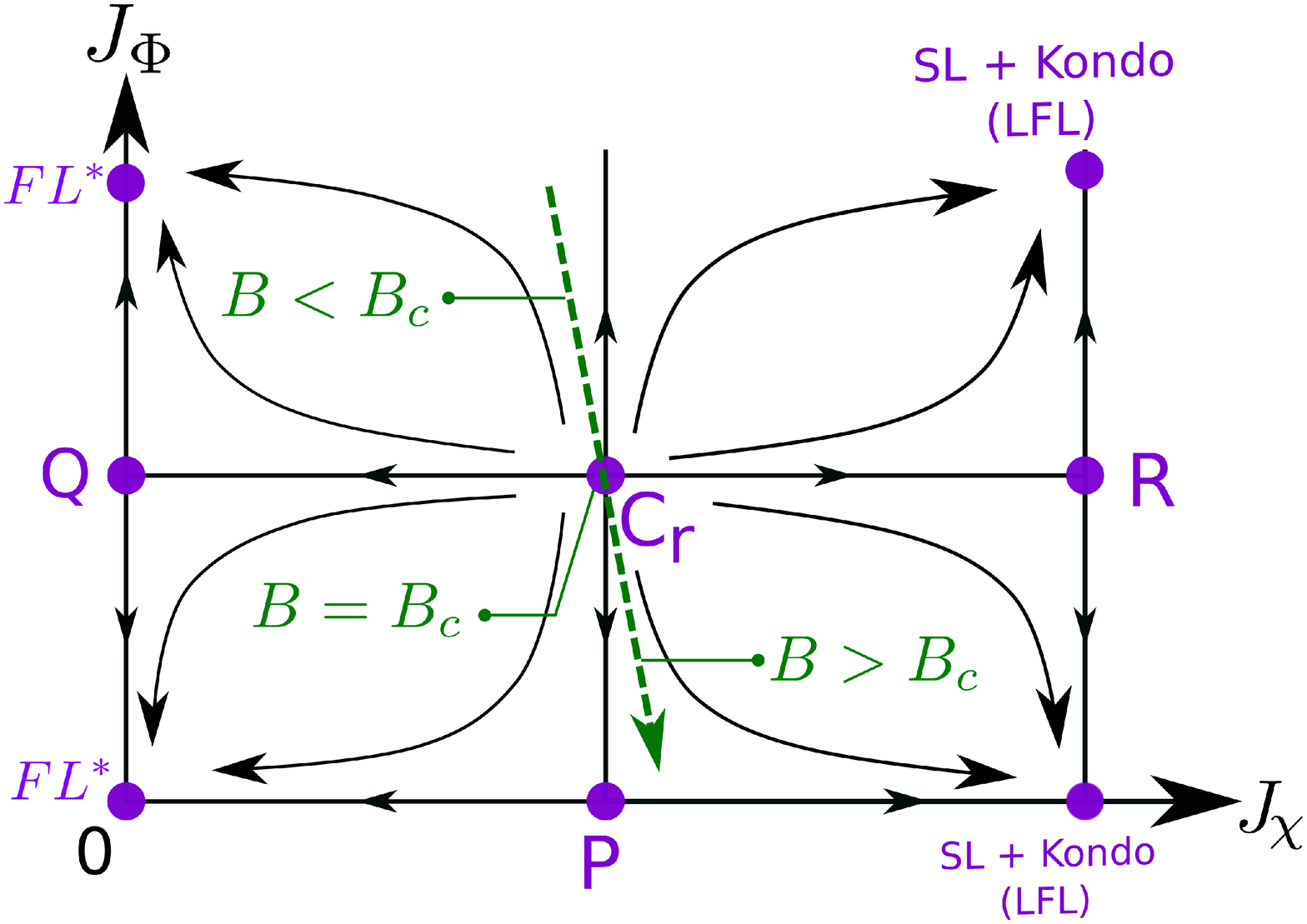}
                \caption{The schematic RG flow diagram shows the competition between $J_{\chi}$ and $J_{\Phi}$ at fixed $u_{\chi} = u_{\Phi} = 0$. The magnetic field in experiments is proposed to follow the green dashed line.} 
                \label{fig:rgflow}
           \end{figure}   
   
{\it The Large-$N$ Mean-Field Hamiltonian}. 
Our starting point is the fermionic $Sp(N)$ large-$N$ mean-field Hamiltonian of the Kondo lattice model \cite{Senthil2003}: 
        $H_{Sp(N)}^{MF}=H_{0}+H_{\lambda}+H_{K}+H_{J}$,
where    
       $H_{0}=  \sum_{\langle i, j \rangle; \sigma}\left[ t_{ij}c^{\dagger}_{i\sigma}c_{j\sigma}+h.c. \right] - \sum_{i\sigma} \, \mu \, c_{i\sigma}^{\dagger} c_{i\sigma}$ , 
       $H_{\lambda}  =  \sum_{i, \sigma}\lambda \left [f_{i\sigma}^{\dagger}f_{i \sigma}-2S \right]$, 
          $H_{J} =  \sum_{\langle i,j \rangle}J_{ij}{\bf S}_{i}^{imp} \cdot {\bf S}_{j}^{imp}
      =  \sum_{\langle i,j \rangle;\alpha, \beta } \left[ \Phi_{ij}\mathcal{J}^{\alpha \beta} f_{i\alpha}f_{j\beta}+h.c. \right]+ \sum_{\langle i,j \rangle}N \frac{|\Phi_{ij}|^{2}}{J_{H}}$,
      $H_{K} =  J_{K} \sum_{i} {\bf S}_{i}^{imp} \cdot {\bf s}^{c} 
=  \sum_{i,~\sigma} 
\left[ \left( c_{i \sigma}^{\dagger} f_{i \sigma}  \right) \chi_{i} +h.c. \right] + \sum_{i}\, N\frac{|\chi_{i} |^{2}}{J_K}$,
where the antiferromagnetic RKKY interaction $H_{J}$ is described by 
 the fermionic $Sp(N)$ spin-singlet with $\left[ Sp(1) \simeq SU(2) ~\text{for}~N=2  \right]$, $\mathcal{J}^{\alpha \beta}=\mathcal{J}_{ \alpha\beta}=-\mathcal{J}^{\beta\alpha }$ being the generalization of the $SU(2)$ antisymmetric tensor $\epsilon_{\alpha \beta}=\epsilon^{\alpha \beta}=-\epsilon_{ \beta\alpha}=i\sigma_{2}$ \cite{ReadSachdev1991, SachdevKagome, Chung2001,Flint2008Nature}. $H_{0}$ describes hopping of conduction $c$-electrons, while $H_{K}$ denotes Kondo interaction.
We assume a uniform $J_{ij}=J_H$ RKKY 
coupling on a lattice with $i,~j$ being nearest-neighbour sites, 
$\sigma,\alpha,\beta \in \lbrace -\frac{N}{2}, \cdots, \frac{N}{2} \rbrace$ with $N\to \infty$. 
$H_{\lambda}$ describes the local impurity $f_{i \sigma}$ electrons with $\lambda$ being the Lagrange multiplier to impose the local constraint 
$\langle \sum_\sigma f_{i\sigma}^{\dagger}f_{i \sigma} \rangle = N\kappa$   
 where a constant $\kappa \equiv 2S$ 
ensures the fully screened Kondo effect \cite{ColemanSchwingerboson,foot-kappa}. 
The mean-field Kondo hybridization and RVB spin-singlet are defined as 
$ \chi_{i} \equiv \langle \frac{J_K}{N}\sum_{\sigma} f^{\dagger}_{i\sigma}c_{\sigma}\rangle$ and $ \Phi_{ij} \equiv \langle \frac{J_{H}}{N} \sum_{\alpha, \beta} \, \mathcal{J}_{\alpha \beta}f^{\alpha \dagger}_{i}f^{\beta \dagger}_{j}\rangle$, respectively \cite{foot-SU2-gauge}. 
 Besides the RVB phase, this approach enables us to describe both the Kondo LFL and the superconducting 
phases \cite{scYRS2016} via Bose-condensing $\chi$-field and the fermionic $Sp(N)$ singlets, respectively \cite{SpNfootnote}. 
Excluding superconductivity here (as it is likely suppressed by magnetic field), 
$H_{Sp(N)}^{MF}$ on a $ 2d$ lattice is shown to have a fractionalized Fermi-liquid 
(FL$^{\ast}$) phase \cite{Senthil2003} with $\chi_i=0$, $\Phi_{ij}\neq 0$ for $J_K\ll J_{H}$,  
and a Kondo-RVB spin-liquid co-existing heavy-Fermi-liquid (LFL) phase with $\chi_i\neq \Phi_{ij}\neq 0$ for $J_K \sim \mathcal{O}(J_{H})$ 
and a large Fermi surface due to Kondo hybridization \cite{Senthil2003,LFLfootnote}. The mean-field result captures qualitatively 
the competition between RVB singlets 
and the Kondo correlations 
near criticality (see Fig.~1(b)). 
However, to account for the NFL and crossovers,  
we shall analyze the dynamics and fluctuations beyond mean-field level  
via a perturbative renormalization group (RG) approach. \\
{\it The Effective Field Theory Beyond $Sp(N)$ Mean-Field.} 
We consider here the Gaussian (amplitude) fluctuations around the mean-field 
order parameters $\chi_{i}$ and $\Phi_{ij}$ with dynamics 
in the FL$^{\ast}$ phase close to the QCP \cite{foot-Higgs,supp}. The effective action $S_{eff}$ for a fixed $N=1$ reads:     
  \begin{widetext}
      \begin{eqnarray}
     S_{eff}&=&S_{0}+S_{\chi} + S_{\Phi}+S_{K}+ S_{J}+S_{G} +S_{4}, \nonumber \\
             S_{0}&= & \int dk \sum_{\sigma = \uparrow \downarrow}\, \  c_{k \sigma}^{\dagger} \, \left( -i \omega +\epsilon_{c}({\bf k}) \right) c_{k\sigma}+f^{\dagger}_{k\sigma} \, \left( - \frac{i \omega}{\Gamma} + \lambda \right) f_{k\sigma},~ S_{\chi} = \int \, dk \sum_{\sigma = \uparrow \downarrow} \left[\chi_{\textbf{k}}  f^{\dagger}_{k \sigma} c_{k\sigma} + h.c. \right] + \sum_{i} \int\,d\tau  |\chi_{i}|^{2}/J_{K},
        \nn
          S_{\Phi} &= &\int \, dk \sum_{\alpha \beta } \left[\Phi_{\textbf{k}} \epsilon_{\alpha\beta} f^{\alpha}_{k }f^{\beta}_{-k}  + h.c. \right] + \sum_{\langle i,j \rangle} \int \, d\tau |\Phi_{ij}|^{2}/J_{H} ,~~  S_{K} =  J_{\chi}\sum_{\sigma=\uparrow\downarrow}\int dk dk^{\prime} \left[(c^{\dagger}_{k\sigma}f_{k^{\prime}\sigma})\hat{\tilde{\chi}}_{k+k^{\prime}}^{\dagger} +h.c.\right],
\nn
S_{J} &=& J_{\Phi} \sum_{\alpha,\beta =  \uparrow \downarrow } \int dk dk^{\prime} \left[\epsilon_{\alpha \beta}\hat{\tilde{\Phi}}_{k} f_{k^{\prime}}^{ \beta}f_{k+k^{\prime}}^{\alpha}+h.c.\right], ~~ S_G =\int \, dk \left[\hat{\tilde{\chi}}^{\dagger}_{k}\, \left( -i \omega +\epsilon_{\chi}({\bf k})+m_{\chi} \right) \hat{\tilde{\chi}}_{k}+ \hat{\tilde{\Phi}}^{\dagger}_{k}\, \left( -i \omega + \epsilon_{\Phi}({\bf k})+ m_{\Phi} \right) \hat{\tilde{\Phi}}_{k}\right] \nonumber \nn
S_{4} &= & \frac{u_{\chi}}{2} \, \int\, d{k_{1}}d{k_{2}}d{k_{3}} \hat{\tilde{\chi}}_{k_{1}}^{\dagger} \hat{\tilde{\chi}}_{k_{2}}^{\dagger} \hat{\tilde{\chi}}_{k_{3}} \hat{\tilde{\chi}}_{-k_{1}-k_{2}-k_{3}}   
				    +  \frac{u_{\Phi}}{2} \, \int\, d{k_{1}}d{k_{2}}d{k_{3}} \hat{\tilde{\Phi}}_{k_{1}}^{\dagger} \hat{\tilde{\Phi}}_{k_{2}}^{\dagger} \hat{\tilde{\Phi}}_{k_{3}} \hat{\tilde{\Phi}}_{-k_{1}-k_{2}-k_{3}} ,
\label{eq:fulllagrangian}
\end{eqnarray} 
\end{widetext}
   with $k= (\omega, {\bf k})$ and $dk = d^{d}{
   \bf k}d\omega$ and $\tau$ the imaginary time.  The actions $S_{\chi}$ ($S_K$) and $S_{\Phi}$ ($S_J$) represents the Kondo hybridization and RKKY interaction at (beyond) the mean-field level, respectively, while $S_G$ ($S_4$) represents the action of the quadratic Gaussian (quartic) fluctuating fields, respectively. 
   $\Phi_{\textbf{k}}$ ($\hat{\tilde{\Phi}}_{k}$) 
   and $\chi_{\textbf{k}}$ ($\hat{\tilde{\chi}}_{k}$) are the Fourier-transformed 
   mean-field variables (amplitude fluctuating fields above mean-field) $\Phi_{ij}$ ($\hat{\Phi}_{ij}\equiv \frac{1}{N} \Sigma_{\alpha\beta}f^{\alpha\dagger}_i f^{\beta\dagger}_j$) and $\chi_{i}$ ($\hat{\chi}_i\equiv \frac{1}{N}\Sigma_{\sigma}f_{i\sigma}^{\dagger} c_{i\sigma}$), respectively.
$\epsilon_{c}({\bf k})$, $\epsilon_{\chi}({\bf k})$ and $\epsilon_{\Phi}({\bf k})$ are the quadratically dispersed kinetic energies of the itinerant electrons $c_{k}$, local singlet $\hat{\tilde{\Phi}}_{k}$ and the Kondo hybridization $\hat{\tilde{\chi}}_{k}$, respectively. 
The quadratic forms of $\epsilon_{\chi}({\bf k})$ and $\epsilon_{\Phi}({\bf k})$ are derived via integrating out $c$-electrons away from Fermi surface \cite{foot-U1,supp}.  
We find  $\hat{\tilde{\chi}}$-field is not Landau damped   
since the imaginary part of its self-energy 
vanishes: $\text{Im} \Sigma_\chi (\omega) \propto \int \, d\epsilon \delta(\omega + \epsilon -\bar{\lambda}) = 0$ as $\bar{\lambda} \gg \epsilon +\omega$ \cite{supp}, leading to a jump in the Fermi volume at the critical point, consistent with experiments \cite{PaschenNat2004, FriedemannPNAS}. \\

{\it RG Analysis:} 
Within our RG scheme $k$, $k_F$ and $\omega$ are 
rescaled as: $k^{\prime}=e^{l}k;~~k_F^{\prime}=e^{l}k_F;~~\omega^{\prime}=e^{zl}\omega$ where the dynamical exponent $z$ is set to $2$. 
 This scheme, distinct from the conventional one \cite{Yamamoto2010,ShankarRG}, allows the Fermi momentum $k_{F}$ to flow the same way as the momentum variable $k$: $[k]=[k_F]=1$, effectively capturing the mixture of electron population in small and large Fermi surface (or the continuous evolution of the Hall coefficient) at finite temperatures \cite{ShankarRG,YeCuO1991, PatchRGfootnote}.
At tree-level, $J_{\chi}$, $J_{\Phi}$, $u_{\chi}$ and $u_{\Phi}$ are irrelevant couplings for $d>z=2$, while $J_{\chi}$, $u_{\chi}$ and $u_{\Phi}$ become marginal for $d=z=2$ ($[J_\chi]=(z-d)/2,[u_\chi]=[u_\Phi]=z-d$), which allows a controlled perturbative RG analysis on the effective action on a quasi-$2d$ lattice: $d=2+\eta$ with $\eta=0^+$ \cite{foot-quasi2d}.\\

The RG $\beta$-functions in the weak-coupling limit, $J_\chi=J_\Phi\to 0$, are readily obtained via diagrammatic perturbative approaches, which include  coupling constants renormalization and the field (or the Green's functions)-renormalization \cite{Zhu2002,Si2003} (see Ref.~\onlinecite{supp}) :
 \begin{align}
          & \frac{dj_{\chi}}{d l}=-\left( \frac{d-z}{2} \right) j_{\chi}+\frac{1}{2} j_{\chi}^3+2j_{\Phi}^{2}j_{\chi} \nn
          & \frac{dj_{\Phi}}{d l}=-\frac{d}{2} j_{\Phi}+4j_{\Phi}^{3}   ~~;~~ \frac{du_{\Phi}}{dl}=-(d-z) u_{\Phi}- 3 u_{\Phi}^{2} \nn
          &\frac{du_{\chi}}{dl}=-(d-z) u_{\chi} - 4 j_{\chi}^{2}u_{\chi} - 3 u_{\chi}^{2} \nn
          &    \frac{d}{dl}\left( \frac{1}{\Gamma} \right)= -z \left( \frac{1}{\Gamma}\right) + 4 j_{\Phi}^{2}  ~~;~~  \frac{dm_{\chi}}{dl}=z m_{\chi}+ j_\chi^{2} \nn
          & \frac{dm_{\Phi}}{dl}=z m_{\Phi},
           \label{eq:RG-eq}
        \end{align}  
            where $dl=-d\ln \Lambda$ with $\Lambda$ being the running energy cutoff within momentum-shell RG is used while the constant $\bar{\lambda}$ is the effective chemical potential of the local $f$-electron \cite{supp}. Here, $j_{\chi,\Phi} \, (J_{\chi,\Phi})$ refers to the renormalized (bare) coupling \cite{supp}.
            At two stable phases, the mass term $m_{\chi,\Phi}$ flows to a massive fixed point, $m_{\chi,\Phi}^\ast\to \infty$. Near the QCP, their bare values vanish linearly with distance to criticality: $m_{\chi,\Phi}\propto |g-g_c|$. 
            Since the effective dimension $d+z > 4$, greater than the upper critical dimension, 
            the Gaussian fixed point $u^\ast=v^\ast=0$ is stable \cite{Ginzburg-footnote}, and violation of hyperscaling is expected \cite{LJZhu}.        
Two non-trivial intermediate critical fixed points 
are found at $(j^{2^{*}}_{\chi}, \, j^{2^{*}}_{\Phi}) = \left( \eta, 0 \right)$ ($P$) and $\left( 0,  \, d/8 \right)$ ($Q$) (see Fig. \ref{fig:rgflow}). The fixed point  
at $Q$ controls the transition between 
the two FL$^*$ fixed points, while  
$P$ 
separates FL$^*$ phase at $j_\chi =0$  from the Kondo co-existing with spin-liquid (LFL) phase at 
$j_\chi \to \infty$.     
           
      To more precisely locate the QCP at finite values of $j_\Phi, ~j_\chi$, the RG equations for $j_\chi, j_\phi$ are obtained near 
      the fixed points $P$ (with $J_\chi$ being fixed at $J_\chi^{*}$) 
      and $Q$ (with $J_\Phi$ being fixed at $j_\Phi^{*}$) (see Ref. \onlinecite{supp}): 
       		$\frac{d\tilde{j}_\Phi}{dl} = \left(	-\frac{d}{2} + \frac{\eta}{2}\right)\, \tilde{j}_\Phi + 4 \, \tilde{j}_\Phi^3,  \,
       		\frac{d\tilde{j}_\chi}{dl} = -(\frac{\eta}{2}+d/4-2(j_\Phi^*)^2)\,\tilde{j}_\chi +\frac{1}{2} \tilde{j}_{\chi}^3.$
      The critical point $C_r$ in Fig. \ref{fig:rgflow}, which controls the FL$^{\ast}$-LFL QPT, is located at the intersect of the above two RG flows:  
      $((j_\chi^\ast)^2, \, (j_\Phi^\ast)^2) = \left( \eta, d/8 \right)$.  Note that $C_r$ is an interacting QCP due to the presence of Kondo interaction. As a result, the $\omega/T$ scaling in dynamical observables is found there via the Kondo breakdown scenario \cite{FriedemannPNAS,supp}.   
     
{\it Critical Properties and Crossovers.}   
 The correlation length $\xi$ diverges near $C_r$: $\xi\sim |g-g_c|^{-\nu}$ with an exponent $\nu$. 
 We find that $\nu$
 is solely determined by the RG flow of $j_\Phi-j_\Phi^\ast$ via $\beta(\tilde{j}_{\Phi})$ \cite{supp}. 
 This yields $\nu=1/z$,  leading to the linear SM-LFL (SM-FL$^\ast$) crossover scale 
 $T_{LFL}$ ($T_{FL^\ast}$) below which $j_\Phi < j_\chi^*$ ($j_\Phi> j_\chi^\ast$):
  $T_{LFL}, T_{FL^\ast}\propto |g-g_c|$, in perfect agreement with the 
  experiment on Ge-substituted YRS \cite{Custers2003, Custers2010}. This suggests that the main effect of magnetic field 
  $B$ in experiment (represented by the coupling $g=g_{0}-J_\Phi/J_\chi$) 
  near the QCP is to suppress $J_{\Phi}$ while $J_\chi$ is near its critical value $J_\chi\sim J_\chi^*$ 
  (see Fig. \ref{fig:rgflow} and Fig.~1(a)). Careful analysis on the pre-factors 
  gives \cite{supp} $\frac{T_{FL^\ast}}{T_{LFL}}\sim (\frac{J_\chi}{|\chi|})^4 \sim L^{-2\eta} \ll 1$ with $L$ being system size, giving rise to the difficulty to observe $T_{FL^*}$ 
  in experiments (see Fig.~1(a)) \cite{foot-FL*}. 
We further identify $T^\ast$ as the crossover scale for the 
onset of the condensate $\chi$ field below which 
$j_\Phi < \chi$. A sub-linear dependence in $g-g_c$ in found: 
$T^* \propto (g-g_c)^{z/d}$ with $z/d \sim 1-\eta/z$, and $\frac{T^\ast}{T_{LFL}}\sim \frac{|\chi|}{J_\chi}>1$, consistent with the experiments \cite{Custers2003, Custers2010}. 
The experimentally observed inverse-in-field divergence in the $A$-coefficient of the $T^2$ term to the resistivity $\rho\sim AT^2$ close to the FL$^*$-LFL QPT is also reproduced: $A\propto \xi^2\sim |g-g_c|^{-2\nu}\sim |g-g_c|^{-1}$. 
The SM region in Ref.~\onlinecite{Custers2010} 
on Ge-substituted YRS 
can be interpreted here as the extended quantum critical regime down to $T_{FL^*}\to 0$. 
 \begin{figure}[ht]
              \centering
                \includegraphics[scale=0.32]{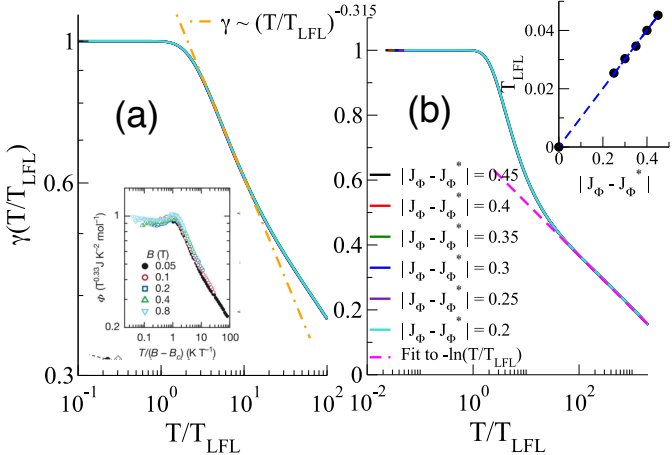}
                \caption{ The log-log (a) and log-linear (b) plots for the ratio $\gamma = C_{V}/\left( T  A(l_0)\right) $  (normalized to $1$ at $T/D_{0}=0$ with $D_{0}$ being the bandwidth cut-off)  at different $|J_{\Phi}-J_{\Phi}^{*}|$ for $\eta=0.18$ 
                with the critical RKKY coupling $J_{\Phi}^{*} = 0.5$ 
                and $J_\chi^* = 0.2$. 
                The inset in (a) shows the experimental data $\Phi(B,T) \propto \gamma (T/T_{LFL})$ 
                taken from Ref.~\onlinecite{Custers2003} for Ge-substituted YRS, inset in (b) shows $T_{LFL}$ v.s. 
                $|J_{\Phi}-J_{\Phi}^\ast|$.  
                Here, we fix the critical Kondo coupling $J_\chi$ and evaluate the specific heat with various $|J_{\Phi}-J_{\Phi}^{*}| = 0.2,~0.25,~0.3,~0.35,~0.4,~0.45.$} 
                \label{fig:Cv}
           \end{figure}

{\it NFL: Electrical Resistivity.} The finite temperature electrical resistivity $\rho(T)$ near the QCP is obtained via the conductivity \cite{HewsonBook}
$\sigma(T)=\rho^{-1}(T)= -\frac{2 e^{2}}{3} \int \frac{d {\bf k}}{(2\pi)^{3}} v_{k}^{2} \, \tau(k) \,\frac{\partial f}{\partial \epsilon_{k}}$ 
with $v_k$ being the electron group velocity, and $\tau^{-1}(\omega)$ the scattering rate of the electron, 
given by the imaginary part of the conduction electron $T-$matrix $\text{Im}\,T(\omega)$ 
: $\tau^{-1}(\omega) = -\frac{c_{imp}}{2}\sum_{{\bf k}} \text{Im} \,T_{{\bf k}{\bf k}}(\omega^{+})$. 
Remarkably, we find that the $T-$matrix contributed from the quasi-2{\it d} bosonic fluctuations in the Kondo hybridization $\hat{\tilde{\chi}}-$field leads to the observed $T$-linear resistivity at low temperatures: $\rho(T) =a^{-1} \, ( 1+c \,T/a)$, where $a$ and $c$ are constant pre-factors \cite{supp}. We find the ratio in resistivity at low temperatures: $\Delta \rho/\rho(T=0) \equiv (\rho(T)-\rho(T=0))/\rho(T=0)$,  
in reasonable agreement with that measured in experiment \cite{Custers2010}.


  {\it NFL: Specific Heat Coefficient.} 
  We further compute the (normalized) scaling function of electronic specific heat coefficient in the SM region: 
  \begin{align}
  	\gamma \left( \bar{T} \right) \equiv \frac{e^{-\eta l}}{T(l) A(l)} \, \frac{\partial \bar{E}_G (l)}{\partial T(l)} \Big |_{l = l_0}=\frac{\bar{T}^{\frac{\eta}{2}}}{4}   \int^{\frac{\Lambda }{\bar{T}}}_{\frac{1}{\bar{T}}} \, dx \, \frac{x^{2+\eta/2}}{\sinh^{2} (x/2)}
  \end{align} (where $\bar{T} \equiv \frac{T}{T_{LFL}} $, $A(l_0)\propto |g-g_c|^{-\zeta}$ with $\zeta \sim 3\eta/2  +\eta^2$ near the QCP,  and  $W(J_\Phi)$ is a non-universal constant), contributed dominantly from the kinetic energy $\bar{E}_G  = \sum_{{\bf k}} \frac{\epsilon_{\Phi}({\bf k})}{e^{\beta \epsilon_{\Phi}({\bf k})}-1}$ of 
  $\hat{\tilde{\Phi}}$ fields \cite{supp}. Here, $l_0$ is a scale at which $m_\Phi (l_0) \sim 1$,
  and $x= \epsilon_\Phi (l)/T(l)$ while $T(l) =  T e^{z l}$ is the scale-dependent dimensionless temperature via the finite-temperature RG scheme \cite{MillisRG}. 
As shown in Fig. \ref{fig:Cv}, $\gamma(\bar{T})$ 
bears a striking similarity to that observed in Ref.~\onlinecite{Custers2003}: it exhibits a power-law scaling behaviour at low 
temperatures before it saturates at $T=0$, i.e. $\gamma(\bar{T}) \propto \bar{T}^{-\alpha}$, 
followed by a logarithmic tail at higher temperatures $\gamma(\bar{T})\propto -\ln \bar{T}$ with  
 exponents $\alpha \sim \zeta \sim 0.3(1)$ for an estimated $\eta\sim 0.18$, in excellent agreement with the experimental values $\alpha\sim \gamma\sim1/3$ \cite{supp}. The $T$-logarithmic behaviour in $\gamma(\bar{T})$ comes as a result of the  Gaussian fixed point for $d=2+\eta$  \cite{foot-uphi}. 
     
{\it NFL: Local Spin Susceptibility.} Finally, the observed anomalous exponent in the divergent 
temperature dependence of the zero-field local spin susceptibility 
$\chi_{\text{loc}}\sim T^{-0.75}\sim\frac{1}{T^{1-\eta_\chi}}$ 
for Ge-substituted YRS  \cite{Vojta2004, Fritz2004, Fritz2006} with $\eta_\chi\sim 0.25$ is reasonably accounted for within 
our approach \cite{supp}: 
              $\chi_{\text{loc}}(T) \propto \frac{J_{\chi}^{4} (T)}{T} 
\propto T^{2\eta-1}$, giving an  
estimated $\eta_\chi\sim 0.36$. 
  {\it Conclusions.} We have theoretically addressed the non-Fermi liquid and quantum critical properties of Ge-substituted YbRh$_2$Si$_2$ by a field-theoretical renormalization group analysis on an effective field theory based on the $Sp(N)$ approach to the Kondo-Heisenberg lattice model. The quantum phase transition and crossover scales are well captured in terms of a competition between a short-ranged fermionic resonant-valence-bond spin liquid and the Kondo effect near a quantum critical point. 
The agreement of our predicted critical properties with experiments is remarkable. 
The strange metal state can be interpreted as the extended quantum critical region 
 to $T\to 0$ due to its proximity to critical Kondo breakdown. 
Our theory 
shed light on the open issues of the non-Fermi liquid behavior in field-tuned quantum critical heavy fermion. 
\acknowledgments
We thank M. Vojta, S. Kirchner, P. Coleman, Q. Si, J. Custers, P. Gegenwart and F. Steglich for helpful discussions. This work is supported by the MOST grant No. 104-2112- M-009 -004 -MY3, the MOE-ATU program, the NCTS of Taiwan, R.O.C. (CHC), and the Austrian Science Fund project FWF P29296-N27 (SP).


\begin{thebibliography}{36}
\bibitem{foot-kappa}
For $N=1$ ($Sp(1)$) and $S=1/2$, $\kappa=1$,  
we find $\lambda\to \infty$. 
In the spin-liquid FL$^\ast$ phase, however, $\kappa <1$ is expected due to lack of magnetic long-ranged order, and $\lambda$ takes a large but 
finite value (see Ref.~\onlinecite{supp}).
\bibitem{foot-SU2-gauge}
For $N=1$, the $SU(2)$ gauge symmetry for the local $f$-electron allows for RVB mean-field variable in the particle-hole channel:  $h_{ij}=\langle \sum_{\sigma=\uparrow,\downarrow} f_{i\sigma}^\dagger f_{j\sigma}\rangle$, and 
 the Kondo hybridization in the particle-particle channel: $\xi_i = \langle\sum_{\sigma}c^\dagger_{i\sigma} f^\dagger_{i\sigma}\rangle$. 
 However, $\xi_i$ can be eliminated by choosing a particular gauge, and $h_{ij}$ is absent due to a different ($Sp(N)$) symmetry exhibited in our model.
\bibitem{foot-U1}
The local $U(1)$ gauge symmetry is broken by $\langle \Phi_{ij}\rangle \neq 0$ boson in both the FL$^\ast$ and LFL phases, leading to both $\hat{\tilde{\chi}}$ and $\hat{\tilde{\Phi}}$ fields to acquire dynamics beyond mean-field.
\bibitem{PatchRGfootnote}
Our RG procedure doe not require to apply the patch RG scheme on a 2d Fermi surface (see Ref.~\onlinecite{supp}).
\bibitem{Ginzburg-footnote}
It can be shown that the ratio $u_{\Phi}(l)/m_{\Phi}(l)$ approaches to zero with the increase of $l$ in the vicinity of the Gaussian fixed points, satisfying the Ginzburg criterion.
\bibitem{foot-FL*}
Further theoretical and experimental studies on the local spin susceptibility at low temperatures near QCP are helpful to clarify the FL$^\ast$ phase.
\bibitem{foot-uphi}
We find a neglibile correction to $\gamma(\bar{T})$ contributed from the quartic $u_{\Phi}$ (see Ref.~\onlinecite{supp}).  
However, these dangerously irrelevant terms may affect critical exponents at ground state and break the hyperscaling relations.
\end{thebibliography}
\end{document}